# An Improved Multiple Faults Reassignment based Recovery in Cluster Computing

Sanjay Bansal, Sanjeev Sharma

**Abstract**—In case of multiple node failures performance becomes very low as compare to single node failure. Failures of nodes in cluster computing can be tolerated by multiple fault tolerant computing. Existing recovery schemes are efficient for single fault but not with multiple faults. Recovery scheme proposed in this paper having two phases; sequentially phase, concurrent phase. In sequentially phase, loads of all working nodes are uniformly and evenly distributed by proposed dynamic rank based and load distribution algorithm. In concurrent phase, loads of all failure nodes as well as new job arrival are assigned equally to all available nodes by just finding the least loaded node among the several nodes by failure nodes job allocation algorithm. Sequential and concurrent executions of algorithms improve the performance as well better resource utilization. Dynamic rank based algorithm for load redistribution works as a sequential restoration algorithm and reassignment algorithm for distribution of failure nodes to least loaded computing nodes works as a concurrent recovery reassignment algorithm. Since load is evenly and uniformly distributed among all available working nodes with less number of iterations, low iterative time and communication overheads hence performance is improved. Dynamic ranking algorithm is low overhead, high convergence algorithm for reassignment of tasks uniformly among all available nodes. Reassignments of failure nodes are done by a low overhead efficient failure job allocation algorithm. Test results to show effectiveness of the proposed scheme are presented.

**Index Terms**— Failure Recovery, Failure Detection, Message Passing Interface (MPI), Redistribution, Load Reassignment

---

## 1 INTRODUCTION

Distributed Computing uses multiple geographically distant computers and solves computationally intensive task efficiently **[1].** There are certain strong reasons that justify using distributed computing in comparison to single powerful computer like mainframes. Cluster computing is one way to perform distributed computing. Several computing nodes connected together form a cluster. Several loosely coupled clusters of workstations are connected together by high speed networks for parallel and distributed applications. Cluster computing offers better price to performance ratio than mainframes. If one machine crashes, the system as a whole can still survive in distributed system. Computing power can be added in small increments in distributed systems. In this way incremental growth can be achieved. Cluster computing has increased in popularity due to greater cost-effectiveness and performance. Recent advancement in processors and interconnection technologies has made clusters more reliable, scalable, and affordable **[2]**.

Fault-tolerance is an important and critical issue in cluster computing. Due to very large size and computation complexity, the chances of fault are more. As the size of clusters increases, mean time to failure decreases. In such a situation, inclusion of fault tolerance is very essential. There are certain areas like air traffic control, railways signaling control, online banking and distributed disaster system high dependability and availability are essential. In absence of sufficient multiple fault tolerance, huge human lives and money could be lost. Hence there is a strong need for improved algorithms for multiple fault tolerance with performance. Performance of a multiple fault tolerance mainly depends on performance of a recovery. Reassignment based recovery with performance is an attractive feature of a cluster computing. There are mainly three types of approaches of fault tolerance in cluster computing. Hardware based fault tolerance is very costly. Software algorithm is only possible when source codes are available. Software layer based overcomes all these problems. In a software layer based fault tolerance, mechanism works as a layer between application and system. This fault tolerance scheme is independent of the cluster scalability as well as fully transparent to the user **[3]**. Several approaches are proposed by researchers. L. Kale and S. Krishnan proposed Charm++ **[4].** Charm++ is a portable, concurrent, object oriented system based on c++. Zheng et al. discussed a minimal replication strategy within Charm++ to save each checkpoint to two "buddy" processors **[5]**. Chakravorty et al. add fault tolerance via task migration to the adaptive MPI system **[6] [7] [8]**. Yuan Tang et al. proposed checkpoint and rollback **[9]**. However, this system's main drawback is expensive in terms of time and overhead. Their system relies on processor virtualization to achieve migration transparency. John Paul Walters et al. proposed replication-Based fault tolerance for MPI application **[10]**. However issues related to replication like consistency among replica and encoding overhead need to be addressed carefully. Number of backups is a major problem in replication based multiple fault tolerance techniques. In order to reduce the number of backups fusion based approach is used **[11].** It is emerging as a popular technique to handle

- *Sanjay Bansal is with theMedi-Caps Institute of Technology, Indore (India).*
- *Sanjeev Sharma is with Rajeev Gandhi Prodyogiki Vishwavidhalaya Bhopal (India).*

multiple faults. Basically it is an alternate idea for fault tolerance that requires fewer backup machines than replication based approaches. In fusion based fault tolerance technique, backup machines are used which is cross products of original computing machines. These backup machines are called as fusions corresponding to the given set of machines **[12]**. Overhead in fusion based techniques is very high during recovery from faults. Hence this technique is acceptable if the probability of fault is low. In this paper we discussed only recovery based on reassignment of tasks to the all the available nodes evenly using a rank based algorithm. Uniform distribution of load among all nodes improves performance of system. Rank based algorithm has low overhead and is efficient since it requires minimal communication among nodes to reassign the task evenly among the nodes.

## 2 CLUSTER COMPUTING

A cluster is a type of parallel or distributed processing system which consists of a collection of interconnected computers cooperatively working together as a single, integrated computing resource **[2]**. Cluster is a multi-computer architecture. Message passing interface (MPI) or pure virtual machine (PVM) is used to facilitate inter process communication in a cluster. Clusters offer fault tolerance and high performance through load sharing. For this reason cluster is attractive in real-time application. When one or more computers fail, available load must be redistributed evenly in order to have fault tolerance with performance. The redistribution is determined by the recovery scheme. The recovery scheme should keep the load as evenly distributed as possible even when the most unfavorable combinations of computers break down **[13].**

## 3 RELATED WORK

Here we review various redistribution based recovery algorithms. Niklas Widell proposed Match-maker algorithm which is pairing based load distributions algorithm. It performs load distribution evenly by pairing overloaded nodes with under-loaded ones, initiating module migration within the pair. The Matchmaker algorithm is found to be fast and efficient in reducing load imbalance in distributed systems, especially in large systems **[14].**

Tree structure is used to redistribute the loads evenly among nodes. Uniform redistribution is done between adjacent nodes if is not a leaf node. If it is a leaf node, it can either balance load with its adjacent nodes or find another leaf node, which is lightly loaded node. Specifically, when a leaf node becomes overloaded, it first tries to do load balancing with its adjacent nodes. If its adjacent nodes are also heavily loaded, then it finds a lightly loaded node for load balancing. Without loss of generality, this lightly loaded node is considered towards the right of the overloaded node. The lightly loaded node can pass its load to right adjacent node. It then leaves the current position in the network and re-joins as a child of the overloaded node, with forced restructuring of the network (to the left for the node leaving and to the right for the node joining) if necessary [15].

Partitioning is one of the approaches by which fast and even load redistribution can be done. In partitions approach, partition is done between the various computing loads in sub domains based on some parameters. Various partitioning methods are suggested by researchers to develop an adaptive and dynamic load distribution. Mohd. Abdur Razzaque and Choong Seon Hong suggested an improved subset partioning of all nodes to reduce the scheduling overhead. This algorithm handles the task of resource management by dividing the nodes of the system into mutually overlapping subsets. Thereby a node gets the system state information by querying only a few nodes. Thus**,** scheduling overheads are minimized **[16].** Another approach is most to least loaded (M2LL) policy. This policy aims at indicating pairs of processors. The M2LL policy fixes the pairs of neighboring processors by selecting in priority the most loaded and the least loaded processor of each neighborhood. Its convergence is very fast. It takes less number of iterations as compare to Relaxed First Order Scheme (RFOS).It also produces more stable balance state **[17].** Binary tree structure is another approach used to partition the simulation region into sub-domains. From a global view to local view, it redistributes the loads between sub-domains recursively by compressing and stretching sub-domains in a group. This method can adjust the sub-domains with heavy loads and decompose their loads very fast **[18].**

In this paper, a new pairing based algorithm is proposed. This algorithm is fast redistribution and balancing algorithm because load is transferred between highest loads to lowest load, second highest to second least and so on. These results in enhanced splitting ratio, hence it causes less number of iterations to converge. While transferring the load, the average load at every node is assured**.** Amount of load transferred is the difference of excess load available at highly loaded node and moderately loaded node so amount of load transfer is also in its permissible value. Another feature of proposed algorithm is that it takes less number of iterations and time to balance the load. Therefore its convergence is fast. It requires fewer efforts to assign loads of failure nodes and new load by just finding the least load node among the several nodes. It also requires less communication overheads since only two nodes have to communicate with each other instead of all. Less number of computer nodes are required to query for load redistribution hence scheduling overheads are minimized.

## 4 PROPOSED ALGORITHM FOR RECOVERY

In proposed recovery scheme, redistribution is done with rank based algorithm. A rank based algorithm is proposed. This algorithm assigns a rank based on their load. After failure detection of failed nodes and recovery of message lost, available loads of all failure and working nodes are redistributed uniformly among all available nodes by this algorithm. This rank based algorithm redistributes the load evenly, quickly and efficiently. This reduces the overall execution time of the system. This algorithm is based on following assumptions and conditions for distributed environment:
(1) Jobs are independent
(2) All jobs of same nature as per communication and computation.
(3) All computing nodes have huge and equal computing capability.
Proposed recovery scheme is composed of two phases
(1) Sequential Phase
(2) Concurrent Phase

### 4.1 Informal and formal Description of the Sequential load redistribution algorithm

In this phase, redistribution of loads of all working nodes is done with a dynamic rank based algorithm. This having following major components in the architecture shown fig. 1:-
(a) Basic rank table generator
(b) Load distributor

### (a) Basic rank table generator

Load information module collects the load of different computing nodes. Based on load information, processors are ranked. Ranking is based on giving the lowest rank to a processor having the least load node and the highest to the node having the highest load. In this way, a rank is allocated to each node based on their load value. A higher rank is allocated when node is heavily loaded. A lower rank means node is lightly loaded. A load is transferred by load transfer module between the highest rank and lowest rank, a second highest ranked node to second lightly ranked node and so on. This module uses rank allocation algorithm shown in Algorithm 1 to generate the Rank table 1 of figure 2.

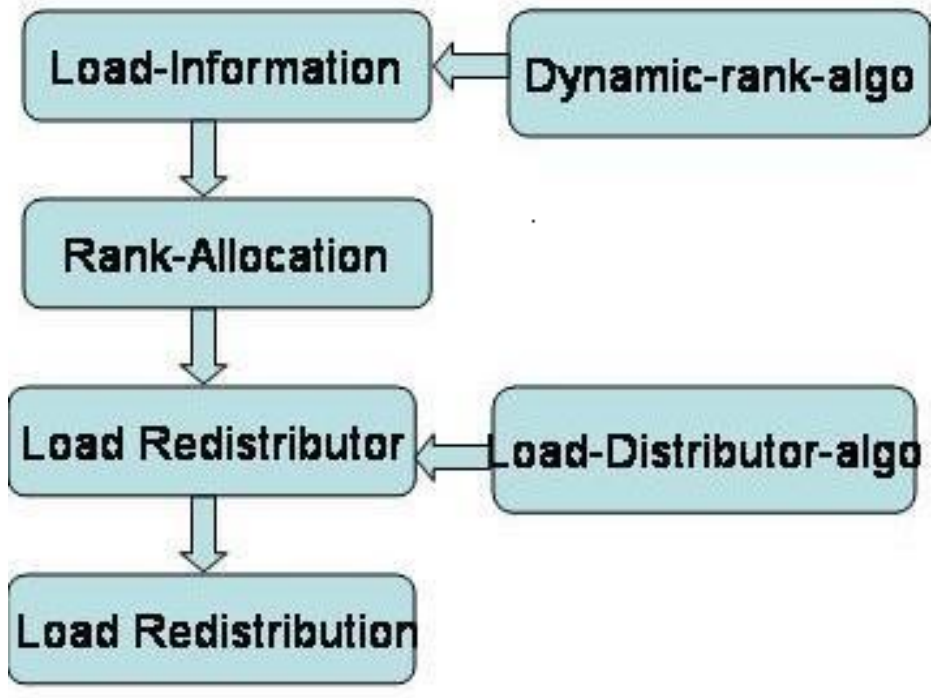

Fig 1. Basic Architecture of Sequential Redistribution

Algorithm1: *Rank allocation*
Begin
1. define an array for storing the load of all nodes $l[]=\{l_1,l_2,l_3,---l_n\}$;
2. define rank an array for storing rank for allocation r[]={1,2,3,4------n};
3. define variable max_load, min_load,max_rank, min_rank and initialize all to zero
4. for j=1 to n do
        if (load[j] > max_load)
    {

      if (i==2)
     {
      make min_rank equal max_rank;
      make min_load equal to max_load;
     }
    assign rank of j node to one value more than max_rank;
    set value of max_rank to rank of j;
    max_load=load[j];
    change the value of max_load to load value of j;
    increase i by 1;
    if (load[j] < min_load)
    {
    assign rank of j node to one less than min_rank;
    set the min_rank value to rank of j node;
    set the value of min_load to load of j node;
   }
   if (load[j] > min_load && load[j] <max_load)
   {
    set rank of j node to one more value of min_rank;
     }
   }
5. End.

This algorithm works sequentially and generates the rank table 1 of fig 2 given below

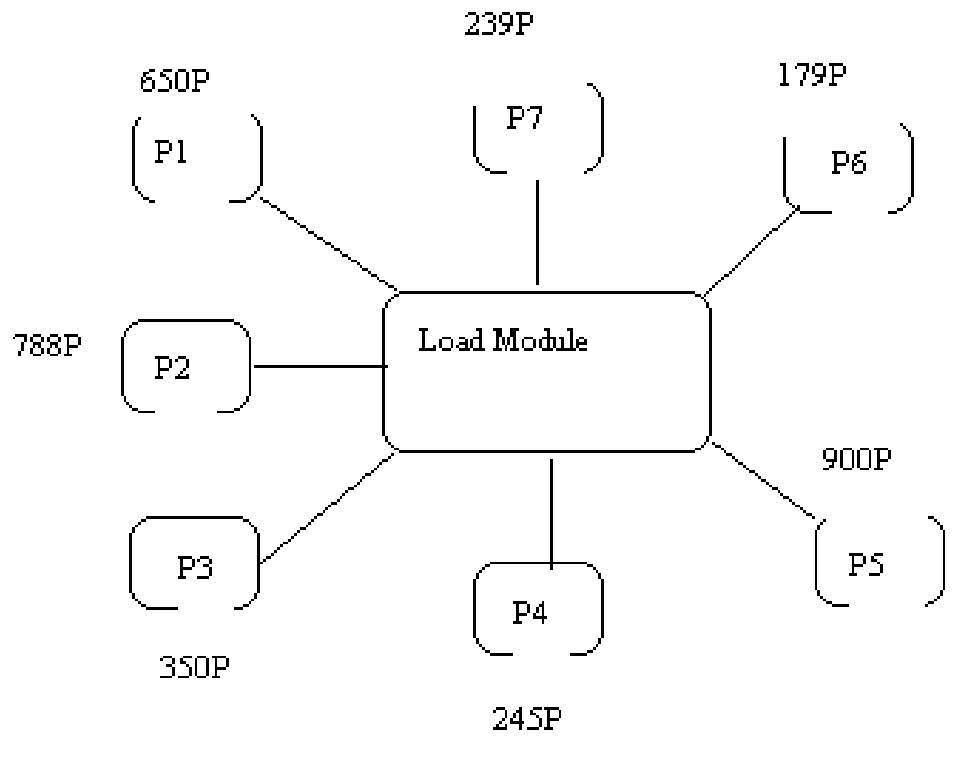

Fig 2. Load module connected to different nodes working in distributed computing system.

**Table 1: Rank Table generated for figure 2 by rank algorithm 1**

| CPU-ID | LOAD | RANK |
|---|---|---|
| P1 | 650 | 5 |
| P2 | 788 | 6 |
| P3 | 350 | 4 |
| P4 | 245 | 3 |
| P5 | 900 | 7 |
| P6 | 137 | 1 |
| P7 | 239 | 2 |

### (b) Load Distributor

Once Rank table load is prepared by the rank generator; the load distributor distributes the loads of all working and non working nodes evenly and uniformly among all the available nodes using load transfer algorithm 2.

**Algorithm 2:** *Load Transfer algorithm*

Load_Redistribution ()
{
1. define variable last to get value of highest rank;
2. define variable rank;
3. define variable avg_load;
4. define var load_to_transfer;
5. retrieve the value of last rank and store it in last';
6. For (i=1;i=last;i++)do

{
avg_load is calculated as avg of i^th^ node and last node;
load_to_transfer is calculated as difference of last node load and avg_load;
transfer load from 1^st^ rank load to current node by load_to_transfer;
assign rank of last node to i;
decrement last variable by 1;
}

By algorithm 2 loads redistribution will take place and it will result in Table 2.

Table 2: uniform load distribution of table 1 by algo.2

| CPU-ID | LOAD |
|---|---|
| P1 | 448 |
| P2 | 426 |
| P3 | 436 |
| P4 | 448 |
| P5 | 426 |
| P6 | 426 |
| P7 | 426 |

Now with final table all nodes are approximate at equal load further load balancing is not suggested

### 4.2 Concurrent redistribution of failure nodes tasks and new tasks

First search the least ranked node and transfer job to it using following algorithm. This algorithm runs concurrently after successfully balanced the loads among all working loads.

Algorithm 3: *failure_nodes_cum_new_job_ assignment*.
Allocate_failure_job ()
{
While (failure_node_job_available () or new_job())
{
Node=get_least_rank_node ();
Compute (Node, job);
}

## 5. Experimental Result and Performance

We have simulated our result on computer nodes of 4 computers. All computers had equal computing power. We had written a process using MPI for printing hello. An experimental result shows reduction in response time due to uniform distribution of tasks of all working and non working node by proposed recovery scheme.

Table 3: Performance comparison after recovery with uniform load distribution

| Load Allocation To All 4 Nodes | Response Time in (ms) | |
|---|---|---|
| | Purposed Method | Simple Mpi |
| 4,9,32,40 | 223 | 640 |
| 1,10,47,50 | 297 | 691 |
| 1,10,59,60 | 310 | 724 |
| 0,0,69,70 | 319 | 988 |

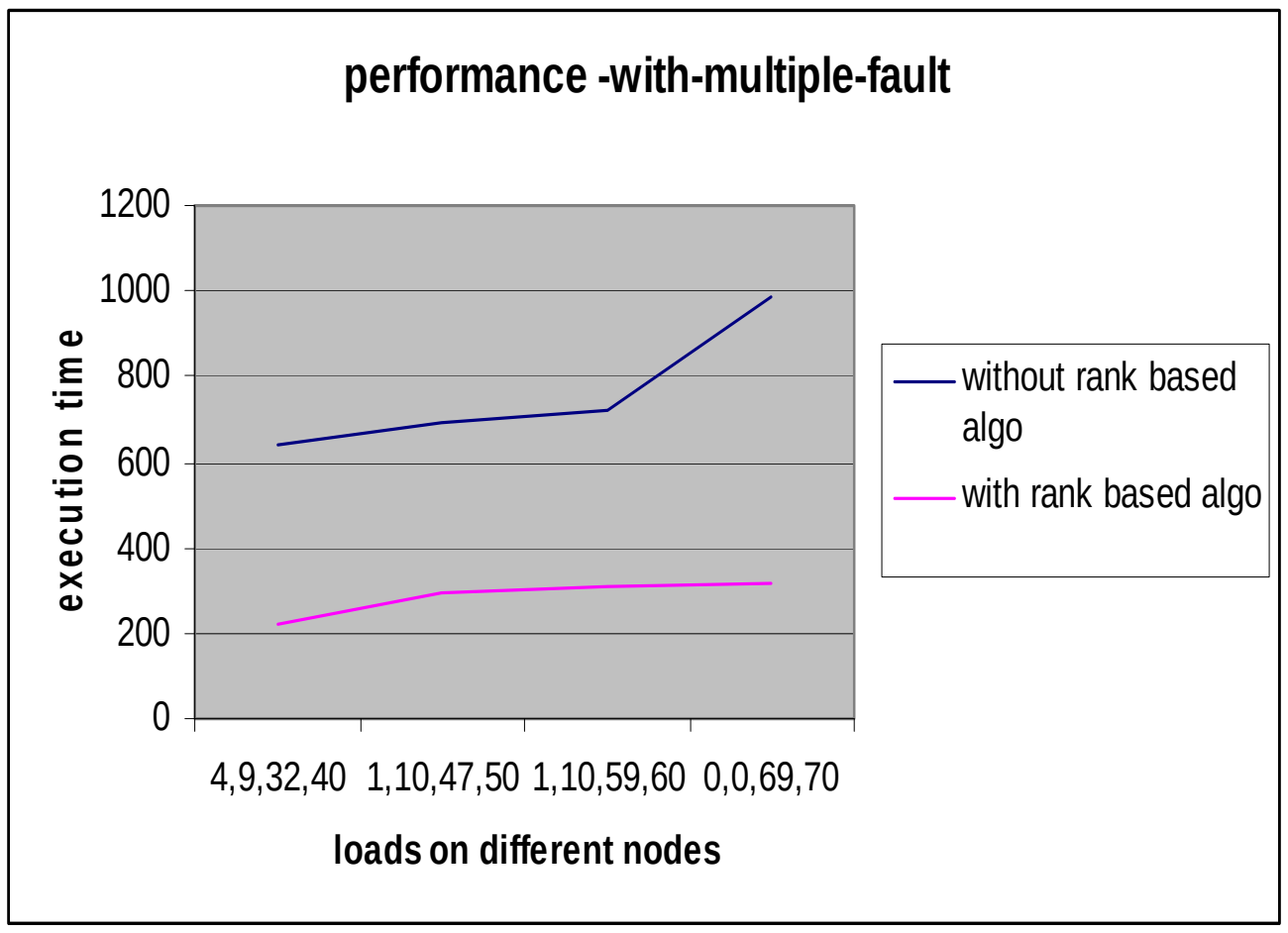

Fig 3: Performance comparison after recovery with uniform load distribution

## 6 Conclusion and Future Scope

Multiple faults tolerance performance depends on accuracy of detection and performance of recovery. Recovery is based on reassignment of tasks. Reassignment is based on distributing the load uniformly among all working nodes. It reduces the response time. Rank based algorithm uses high splitting ratio, hence convergence is very fast. It takes less number of iterations and time. Rank based algorithm is simple and effective. This algorithm having less execution time, fast convergence and fewer messages overhead as compared to other algorithm purposed. This recovery scheme is transparent to user as well. Although this algorithm is suitable for homogenous environment but it can be further extended to heteroge-neous environment as a future enhancement.

**Sanjay Bansal** has passed B.E. (Elex & Telecom Engg.) and M.E. (Computer Engineering) from Shri Govindram Seksariya Institute of Technology and Science, Indore in 1994 and 2001 respectively Presently he is working as Reader in Medi-Caps Institute of Technology, Indore. He is pursuing PhD from Rajeev Gandhi Proudyogiki Vishvavidyalaya, Bhopal, India. His research areas are load balancing, fault-tolerance, performance and scalability of distributed system.

**Sanjeev Sharma** has passed B.E (Electrical Engineering) in 1991 and also passed M.E. in 2000.He is PhD. His research areas are mobile computing, data mining, security and privacy as well as adhoc networks. He has published many research papers in national and international journals. Presently he is working as an Associate-professor in Rajiv Gandhi Proudyogiki Vishwavidyalaya Bhopal (India)